\documentclass[14pt,preprintnumbers,amsmath,amssymb]{article}
\usepackage{graphicx}% Include figure files
\usepackage{dcolumn}% Align table columns on decimal point
\usepackage{bm}% bold math
\author{A. Lohrasebi$^{1}$, M. Neek-Amal$^{2}$, and M. R.
Ejtehadi$^{3}$\\
\small $^1$Department of Physics, University of Isfehan, Isfehan,
 Iran.\\
\small $^2$Department of Physics, Shahid Rajaee Teacher Training
University, Lavizan,
Tehran 16788, Iran.\\
{\small $^3$ Department of Physics, Sharif University of Technology,
Tehran P.O.Box 1155-9161, Iran.}}
\begin{document}
\title{\bf Directed motion of $C_{60}$ on a graphene sheet subjected to
a temperature gradient}
\date{\today}
\maketitle
\begin{abstract}
 Nonequilibrium molecular dynamics simulations is used to study the motion of a $C_{60}$ molecule on a
graphene sheet subjected to a temperature gradient.
 The $C_{60}$
molecule is actuated and moves along the system while it just
randomly dances along the perpendicular direction. Increasing the
temperature gradient increases the directed velocity of $C_{60}$. It
is found that the free energy decreases as the $C_{60}$ molecule
moves toward the cold end. The driving mechanism based on the
temperature gradient suggests the construction of nanoscale
graphene-based motors. \\

\end{abstract}
\maketitle

\section{Introduction}

Since graphene has been discovered~\cite{novoselov}, many properties
of this two-dimensional material have been studied both
experimentally~\cite{geim} and theoretically~\cite{geim2,geim3}.
 In a recent experimental research  the dynamic of light atoms
deposited on a single-layer garphene has been studied by the mean of
 transmission electron microscopy (TEM) technic~\cite{mayer2008}.
The two-dimensional structure of graphene suggests the possibility
of  motion with just two degrees of freedom.  The free energy
surface for a particle moving above a graphene sheet explains
different motion-related phenomena at nanoscale as well as the
various directed motions on the carbon nanotube-based
motors~\cite{neek, science2008}.

A net motion can be obtained from a nanoscale system subjected to a
thermal gradient~\cite{10,11}. Recently the motion of an
experimentally designed nanoscale motor consisting of a capsule-like
carbon nanotube inside a host carbon nanotube has been explained
successfully with molecular dynamics (MD) simulations~\cite{Somada}.
The capsule travels back and forth between both ends of the host
carbon nanotube along the axial direction. Barreiro \emph{et al.}
have designed an artificial nanofabricated motor in which one short
carbon nanotube travels relative to another coaxial carbon
nanotube~\cite{science2008}. This motion is actuated by a thermal
gradient as high as  1\,K\,nm$^{-1}$ applied to the ends of the
coaxial carbon nanotubes.

Since graphene has a very high thermal conductivity
(3000-5000\,W\,K$^{-1}$\,m$^{-1}$~\cite{thermalgraphene,thermalgraphene2}),
as high as diamond and carbon
nanotubes~\cite{thermalcnt,prb2002,prb2004,osman}, it is a good
candidate for heat transferring designs in nano-electromechanical
systems. Because of strong covalent bonds in graphene, thermal
lattice conduction dominates the electrons
contribution~\cite{thermalgraphene}. Recently Yang \emph{et
al.}~\cite{Hu} have studied the thermal conductivity and thermal
rectification of trapezoidal and rectangular graphene nanoribbons
and found a significant thermal rectification effect in asymmetric
graphene ribbons~\cite{Hu}.

Here we study the motion of a nanoscale object, e.g. $C_{60}$, on a
graphene sheet, in the presence of a temperature gradient. We show
that the graphene is a good two-dimensional substrate for thermal
actuation due to its high thermal conductivity, however as it is
expected, in the absence of a thermal gradient, the $C_{60}$
molecule randomly diffuses on the graphene sheet~\cite{neek}. The
average velocity along the temperature gradient direction and the
free energy change throughout the system are calculated.

This paper is organized as follows. In Sec.~2 we will introduce the
atomistic model and the simulation method. Sec.~3
 contains the
main results including those for the produced temperature gradient,
the trajectory of the $C_{60}$ molecule over the graphene sheet and
the free energy change. A brief summary and conclusions are included
in Sec.~4.

\section{The model and method}
\begin{figure}
\begin{center}
\includegraphics[width=0.8\linewidth]{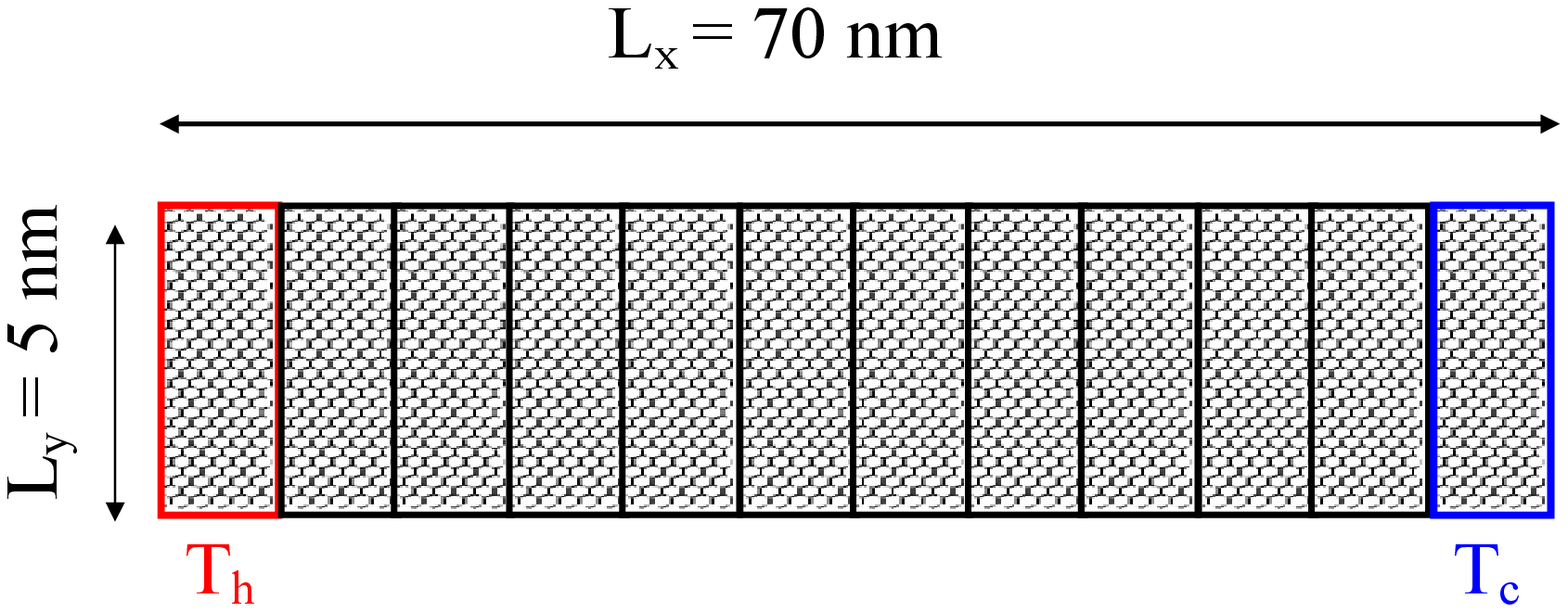}
\includegraphics[width=0.5\linewidth]{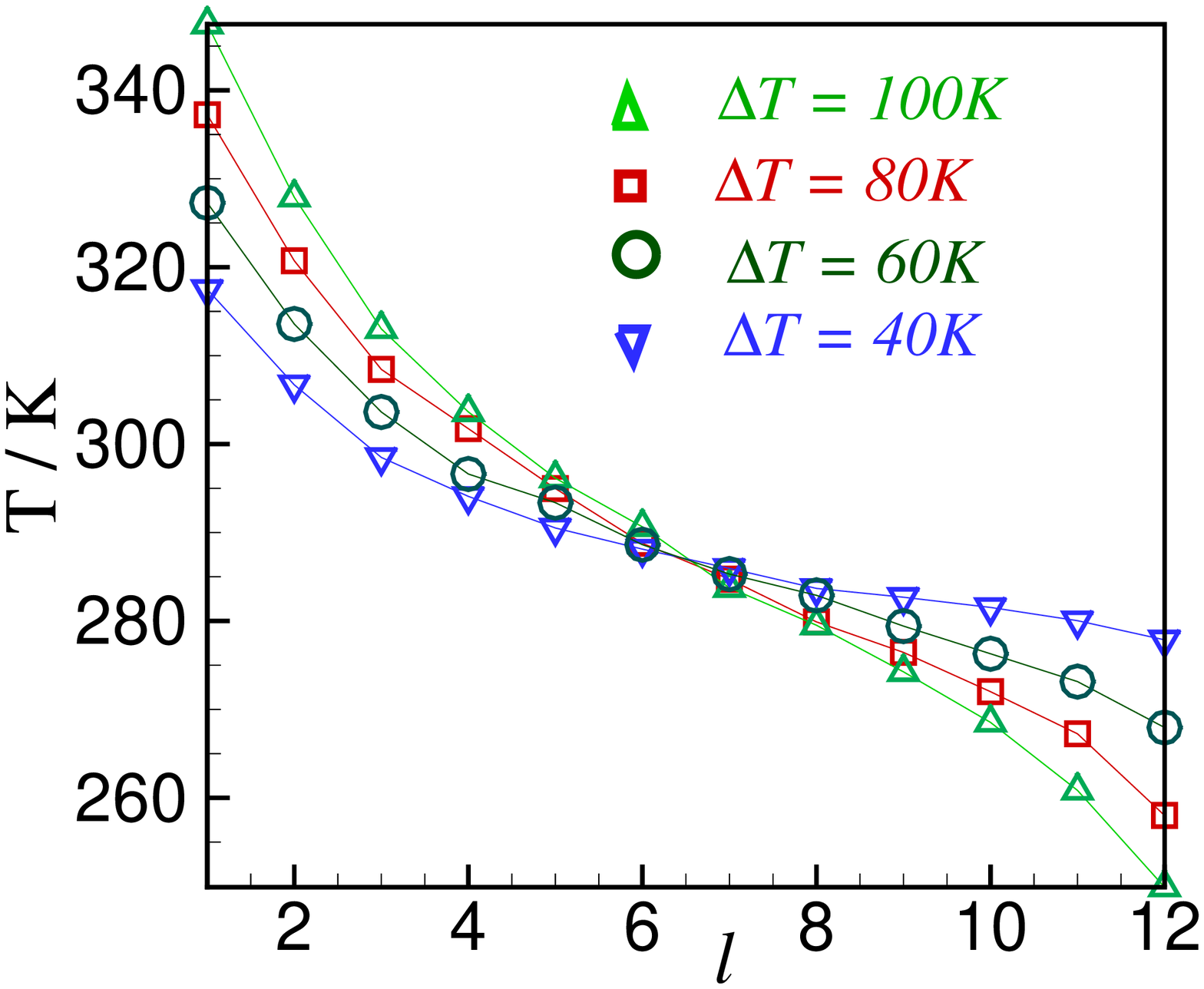}
\caption{(Color online) Top:  The model which shows applied
temperature gradient along $x$-direction. Bottom: Produced four
temperature gradients after 450~ps of a nonequilibrium molecular
dynamics simulation. Delta symbols are related to $\Delta T$~=~100
K, and for square symbols $\Delta T$~=~80 K, circle symbols $\Delta
T$~=~60 K and for gradient symbols $\Delta T$~=~40 K, respectively.
\label{figmode} }
\end{center}
\end{figure}

\begin{figure}
\begin{center}
\includegraphics[width=0.8\linewidth]{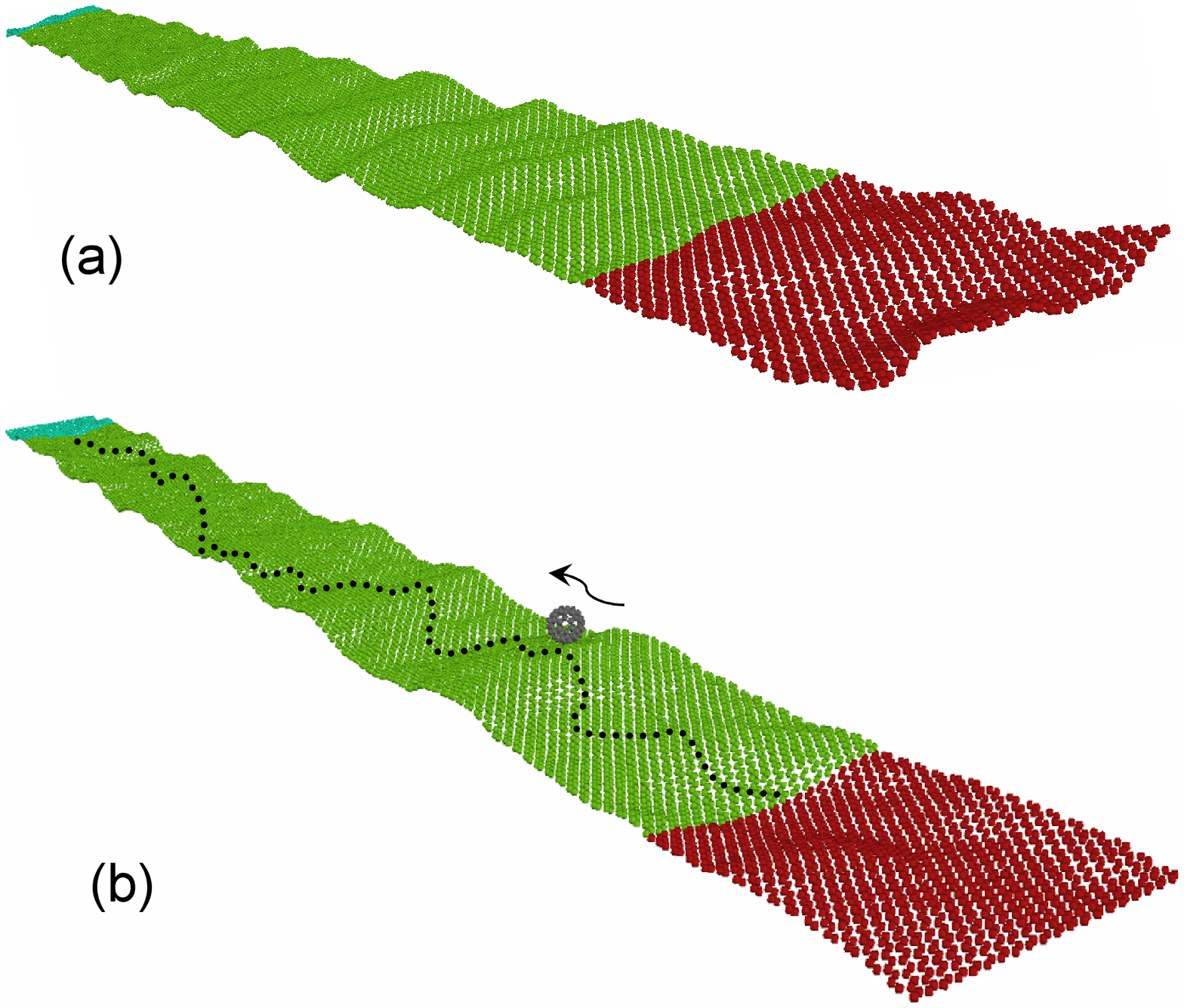}
\caption{(Color online) (a) A snapshot of a not-constrained system
shows the effect of the ends crumpling. (b) Fixing  $z$-components
of the atoms at the first and last rows (see the text) prevents the
crumpling. The black dots show a typical trajectory of the motion of
a C$_{60}$ molecule over the graphene sheet
 from the hot spot
 toward the cold spot.}\label{figcrump}
\end{center}
\end{figure}

The system was composed of a graphene sheet as a substrate, with
dimensions $L_x\times L_y = 70\times$5\,nm$^2$ and a $C_{60}$
molecule above the sheet. The graphene sheet with $N = 14\,400$
carbon atoms was divided into 12 equal rectangular segments. Each
segment with $N_{l} = 1200$ carbon atoms was arranged in 40 atomic
rows (along armchair or $x$-direction). Each row has 30 atoms which
were arranged along zigzag direction. The system was equilibrated
for 300\,ps at $T = 300$\,K before the temperature gradient was
applied. Once the system was equilibrated, the first (hot spot) and
last (cold end) segments of the graphene sheet were kept at $T_{h}$
and $T_{c}$, respectively. A temperature gradient between the two
ends was then produced, i.e. $(T_{h}-T_{c})/L_x$~(top panel of
Fig.~\ref{figmode}).
  To make the model more efficient and prevent
   crumpling (see Fig.~\ref{figcrump}(a)) of the ends, we fixed the
  $z$-components  of the first atomic row in the first
  segment and the last row of the last segment (see Fig.~\ref{figcrump}(b)).

We carried out MD simulations employing two types of the interatomic
potentials: 1) the covalent bonds between the carbon atoms in the
graphene sheet and in the $C_{60}$ molecule are described by Brenner
potential~\cite{brenn} and 2) the non-bonded Van der Waals
interactions between the graphene atoms and those of the $C_{60}$
molecule.  Brenner potential has been parameterized to model $sp^2$
covalent bonds in the graphene, carbon nanotube and $C_{60}$
structures. For the non-bonded potential a Lennard-Jones (LJ)
potential  gives reasonable results~\cite{Hendrick}.
 Here we choose the LJ parameters as  $\epsilon = 2.413$\,meV
 and $\sigma = 3.4$\,\AA~\cite{stan} which represent the depth and range of the LJ potential energy,
respectively. Note that the LJ potential is a simple and commonly
used potential for modeling the interaction between carbon
nanostructures~\cite{Gravil,prlrefereepre}. The equations of motion
were integrated using a velocity-Verlet algorithm with a time step
$\Delta t = 0.5$\,fs. The temperature of the hot end ($T_{h}$) and
the cold end ($T_{c}$) were held constant by a Nos\'{e}-Hoover
thermostat. The temperature of the inner segments were not
controlled by the thermostat. Periodic boundary condition was
applied only in the $y$-direction. Due to the applied temperature
gradient, the system can no longer be described by equilibrium
methods and we shall thus employ nonequilibrium molecular dynamics
simulations. A temperature gradient were produced across the system
during 450\,ps and a stationary state was established. The $C_{60}$
molecule was put above the second segment at $z_{cm} = 8$\,\AA~(here
index $cm$ refer to the center of mass).~During the production runs
of 750\,ps both the $x$ and $y$ positions of the center of mass of
the $C_{60}$ molecule were recorded. Moreover the temperature of
each segment was calculated by measuring the total kinetic energy of
that segment. In our simulations a typical value for the relative
standard deviation of the total energy of the extended system is
about $3.5\times 10^{-5}$. A full simulation run takes about 50\,h
CPU time on a 3.2\,GHz Pentium IV processor with 4\,GB RAM.

In order to compute the change in the free energy, one can employ
the commonly used thermodynamic integration and perturbation methods
~\cite{freeenergyfasolino}. A good estimation for the absolute value
of the free energy requires sampling the whole phase space which  is
not feasible. Jarzynski's method removes this difficulty for
nonequilibrium simulations~\cite{jar}. There is an equality between
the change of the free energy $F$ and the work $W$ applied on the
system (here the $C_{60}$ molecule) ~\cite{jar}
\begin{equation}
\Delta F = -\beta^{-1}\ln\overline{\exp(-\beta W)},\label{jeq}
\end{equation}
where $\beta = 1/k_{B}\,T$ ($T$ is the temperature in each segment),
and the average is taken over different configurations with
different initial conditions. In fact, Eq.~(\ref{jeq}) connects the
change of the free energy (between two equilibrium state) and the
applied work on the system in a nonequilibrium process.

\begin{figure*}
\begin{center}
\includegraphics[width=0.325\linewidth]{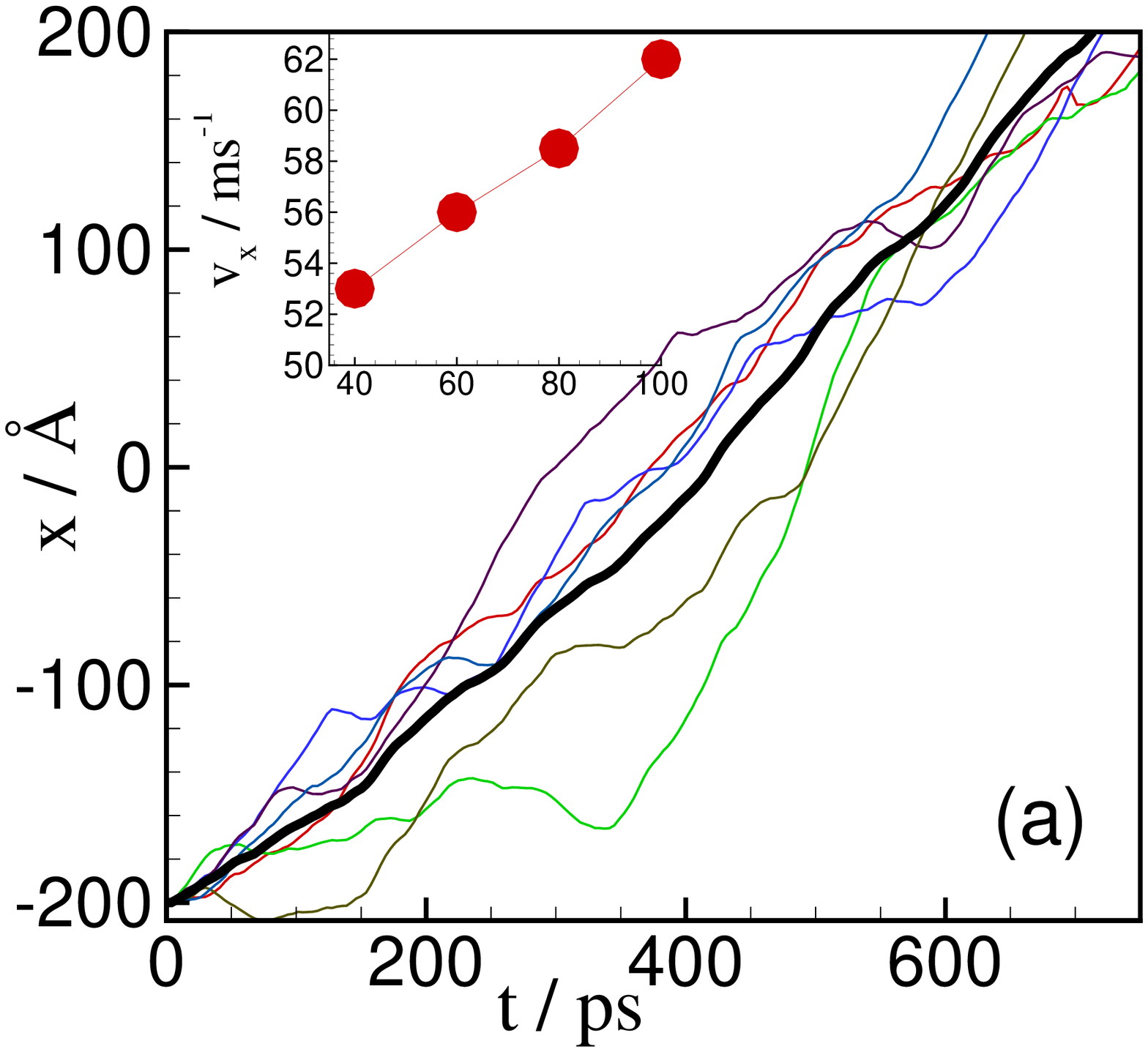}
\includegraphics[width=0.325\linewidth]{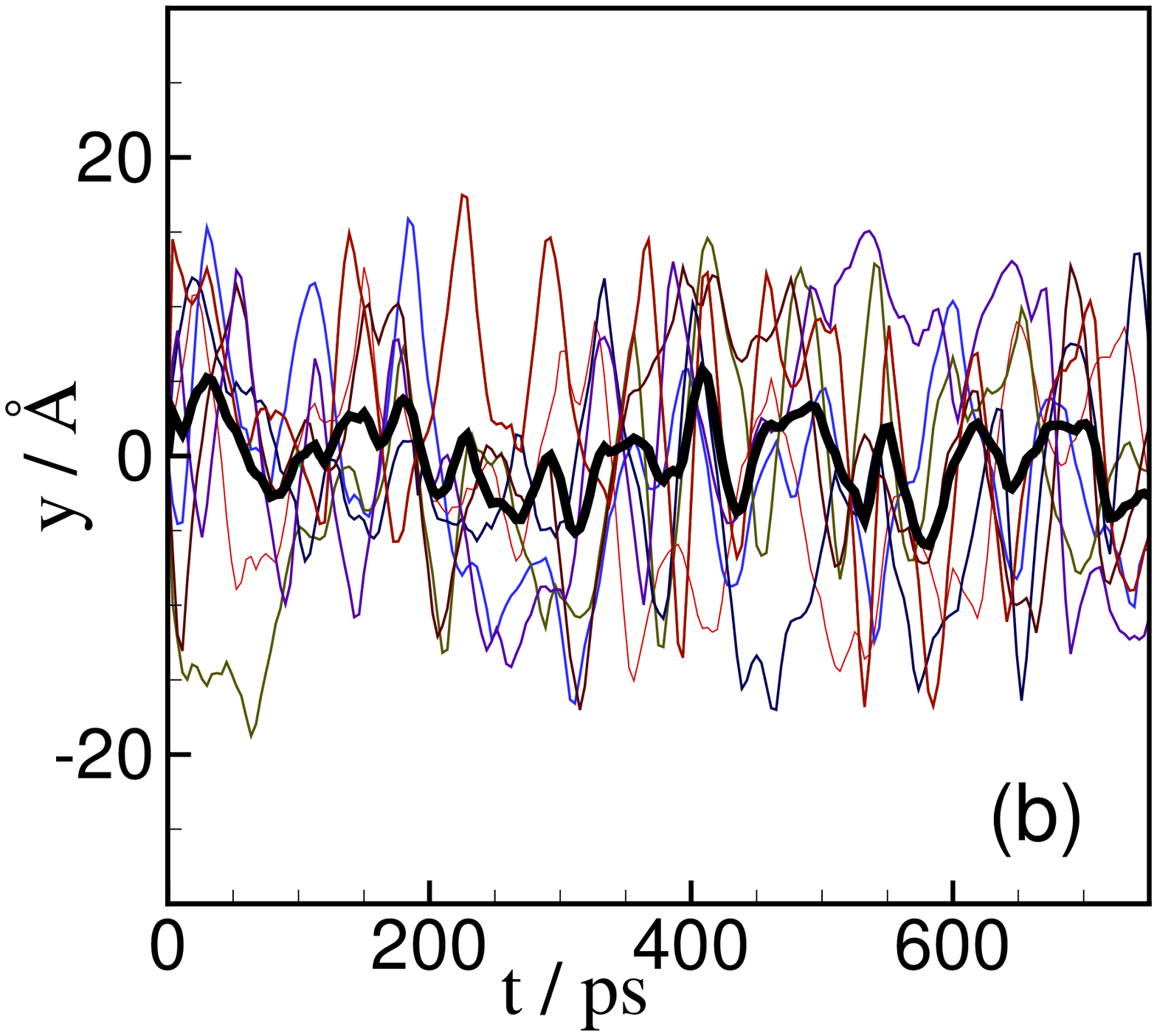}
\includegraphics[width=0.325\linewidth]{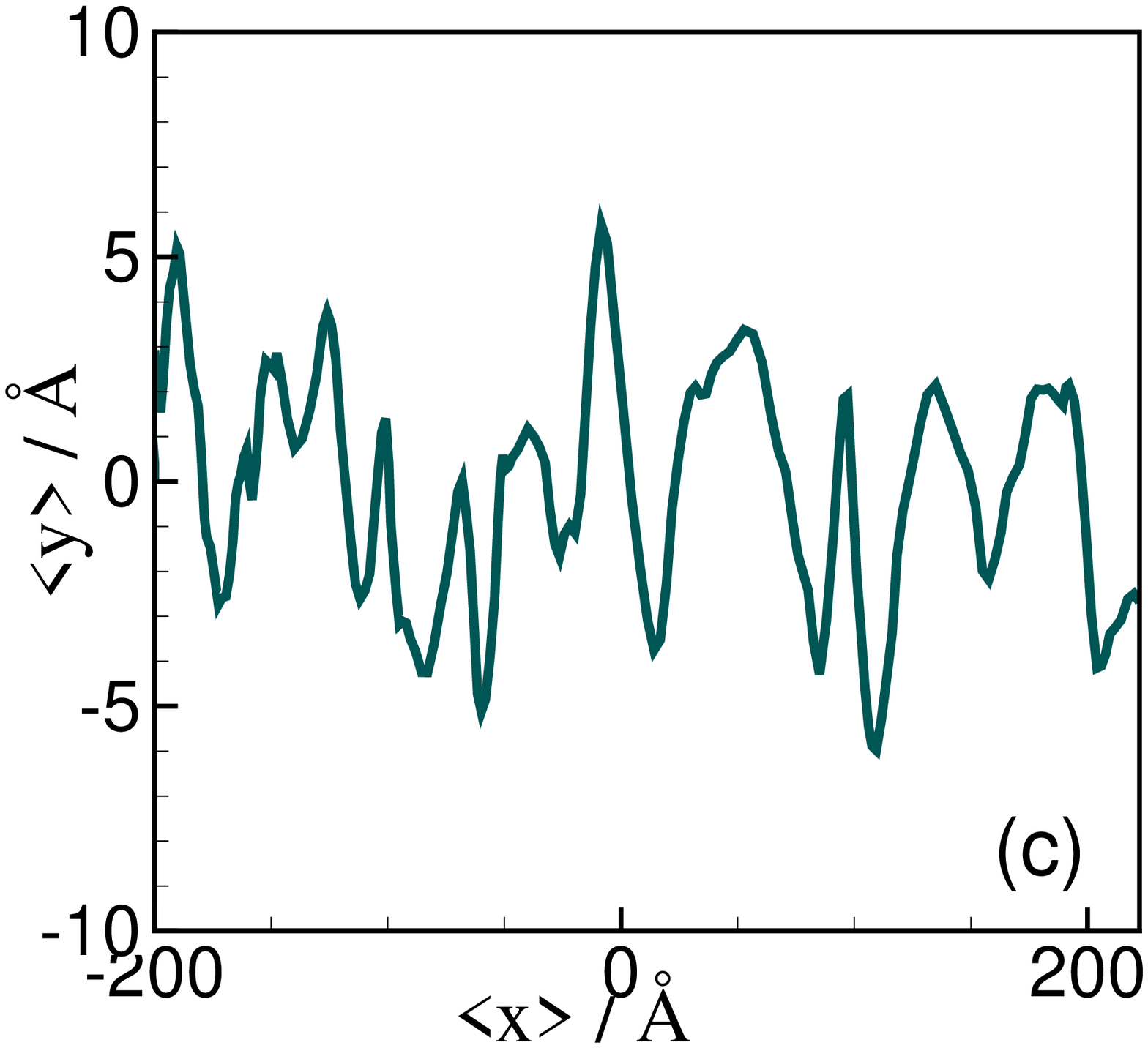}
\caption{(Color online) (a, b) Time series of $x$  and $y$
components of the center of mass of a C$_{60}$ molecule over a
monolayer graphene subjected to a temperature gradient. Thick curves
show the average curves. The inset shows the variation of the
velocity versus temperature gradient. (c) The trajectory of the
motion of a C$_{60}$ molecule (in $x-y$ plane) moving over the
monolayer graphene, averaged over six simulations with different
initial conditions.\label{figxt} }
\end{center}
\end{figure*}

\begin{figure}
\begin{center}
\includegraphics[width=0.5\linewidth]{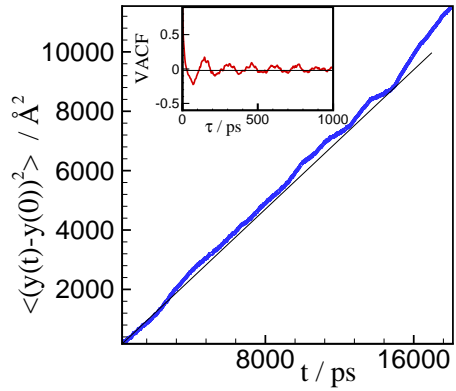}
\caption{(Color online) Mean square displacement for $y$-component
of the center of mass of C$_{60}$ molecule  over the graphene. The
inset shows the velocity autocorrelation function for the
$y$-component of the motion  as a function of time. \label{figMSD} }
\end{center}
\end{figure}

\section{Results and discussion}
\subsection{Producing temperature gradient}

The temperature profiles for different temperature gradients with
$\Delta T =40,~60,~80,~100$\,K are shown in Fig.~\ref{figmode}
(bottom panel). In this figure, the local temperatures of each
segment which were obtained by averaging over 500 data are indicated
by symbols. Corresponding error bars indicate the statistical errors
and are in the range 4-6\,K. Notice that the temperature profiles
are nonlinear which is a commonly observed behavior in
nonequilibrium molecular dynamics simulation of thermal
conductivity~\cite{prb2002,prb2004,osman}. It is a consequence of
the strong phonon scattering caused by the heat source or heat sink
and can be explained by partly diffusive and partly ballistic energy
transport along the system~\cite{prb2002,Oligschleger}. Also,
ripples in the graphene (Fig.~\ref{figcrump}(b)) are the other
important source for phonon scattering~\cite{fasolino,nima}.
Therefore, these mechanisms cause a non-linear temperature profile
in the middle segments.
\begin{figure}
\begin{center}
\includegraphics[width=0.5\linewidth]{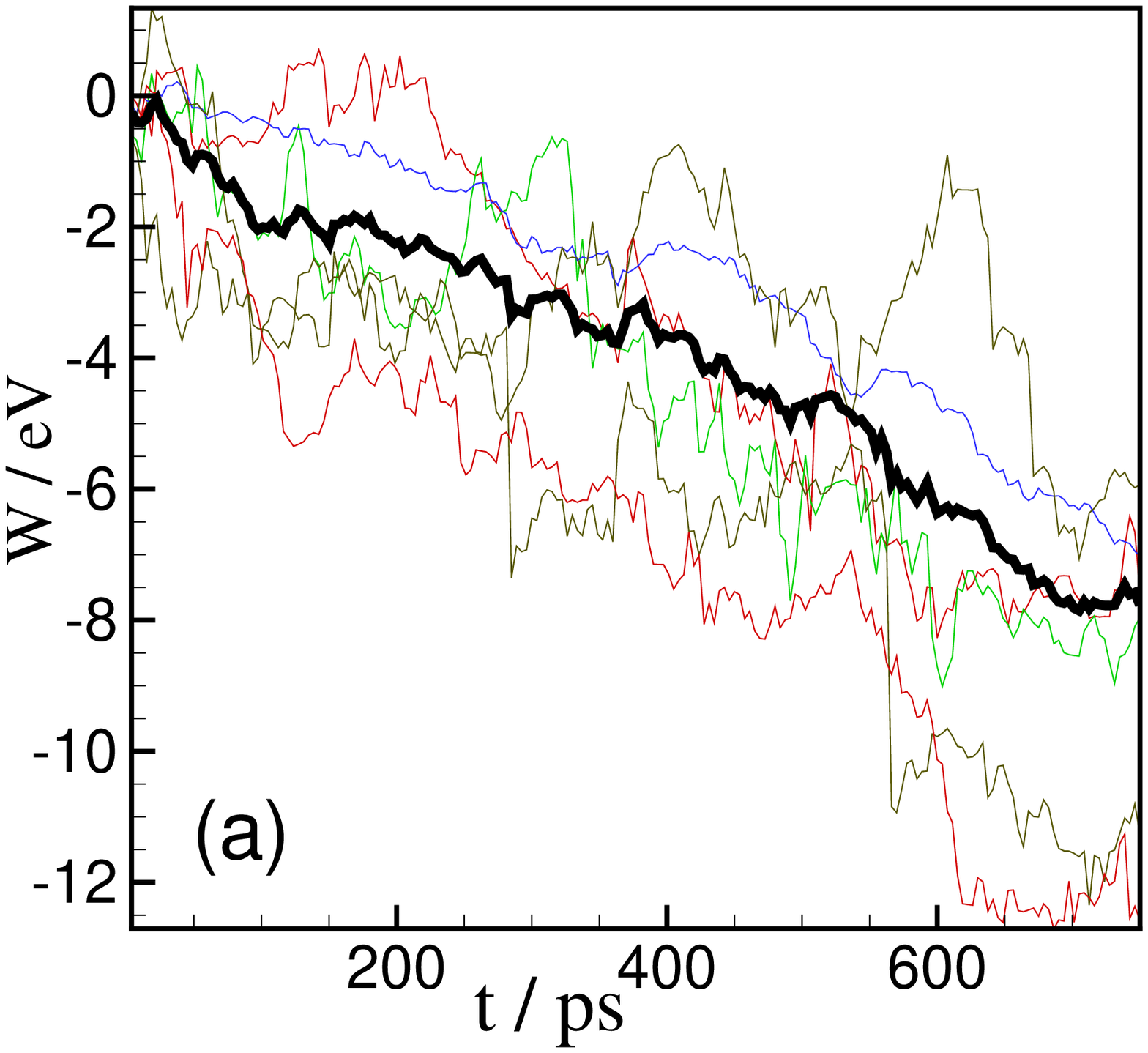}
\includegraphics[width=0.5\linewidth]{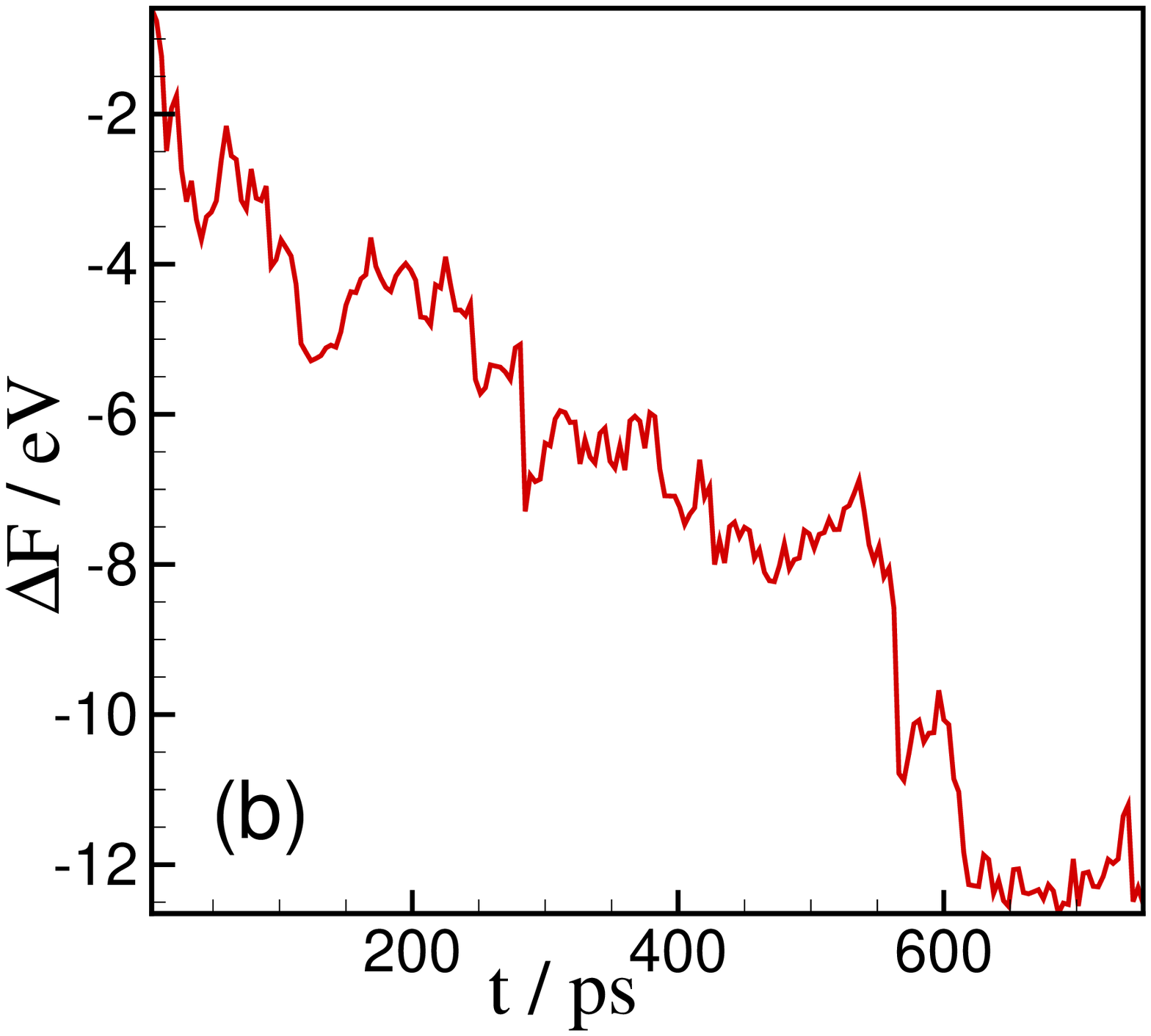}
\caption{(Color online) (a) Total work performed on C$_{60}$
molecules and (b) the change of free energy during C$_{60}$ motion
from the hot end to the cold end. The thick curve is the average
over six simulations with different initial conditions.
\label{figwork} }
\end{center}
\end{figure}

\subsection{Trajectory in the $x-y$ plane}
The graphene ribbon is a two-dimensional way for the motion of the
C$_{60}$ molecule along it. C$_{60}$ attempts to find its
equilibrium state and looks for the local minimum of the free
energy. In other words, in the presence of a temperature gradient,
the phonon waves created in the hot spot travel through the system,
interact and transfer momentum to the C$_{60}$ molecule which
results a net motion~\cite{science2008}.  The time series for the
$x$ and $y$ coordinates, separately, are shown in
Figs.~\ref{figxt}(a,b). The trajectories of the six C$_{60}$
molecules (from the six different simulations) in $x-y$ plane over a
time interval of 750\,ps are depicted in Fig.~\ref{figxt}(c). In all
three cases the thick curve is the average of the six others.

The motion along the $x$-direction is almost a linear uniform motion
with velocities $v_{x} = 53,~56,~59$ and 63\,m/s for $\Delta T =
40,~60,~80,~100$\,K respectively (see inset of Fig.~\ref{figxt}(a)).
There are two computational methods that can be used to  show that
the motion along the $y$-direction is diffusive
(Fig.~\ref{figxt}(c)). First, one can use the Einstein relation to
find the diffusion constant $D_y$ by measuring the slope of the mean
square displacement (MSD) of the $y$-components of the position,
i.e.\ $\langle(y(t)-y(0))^2\rangle$. To show that the motion along
the $y$-direction is driftless, we look at the MSD of $y$-component
of the position of the C$_{60}$ molecule. For a driftless motion the
MSD should grows linearly with time, i.e.
$<(y(t)-y(0))^2>\,=\,2D_yt$. We depict the MSD in Fig.~\ref{figMSD}
for a long time simulation. As we see from this figure  the slope of
the MSD is almost one (after equilibration) which is a signature of
diffusive regime. Alternatively one can use the Green-Kubo relation
which makes use of the velocity autocovariance function (VACF). More
specifically, one can take the integral over the autocovariance of
the $y$-components of the position of the velocity of C$_{60}$ as
$D_y = \int_0^{\infty} \langle v_y(0)v_y(\tau) \rangle d\tau$, which
gives the diffusion constant of random walk motion along the
$y$-axis. The inset of Fig.~\ref{figMSD} shows the VACF versus time.
We have tested both methods in order to compute $D_y$. The result
\textbf for the diffusion constant is about
$4\times10^{-9}$\,m$^2$/s.

 The directed motion
resulted from the temperature gradient along the graphene sheet,
provides a nanoscale motor for the material transferring. Notice
that in the absence of temperature gradient, when the system is
equilibrated at $T$ =~300\,K, we found a diffusive motion on the
graphene sheet~\cite{neek}.

\subsection{Free energy reduction} Figures~\ref{figwork}(a,b) show
the variation of the total work and the changes in the free energy
with time, respectively. The C$_{60}$ loses on average almost 12\,eV
of free energy after 750\,ps during the motion towards the cooler
region. Note that here, the condition $\overline{W}>\Delta F$ is
always satisfied which is a well known criterion for a
nonequilibrium (irreversible) thermodynamics evolution~\cite{jar}.
This method for calculating the change of the free energy is a well
known method in computational soft condensed matter~\cite{hui} but
it turns out to be useful also for studying various physical
properties of graphene, particularly investigating the stability of
new designs for nanoscale molecular devices as studied here.

\section{Conclusions}
In summary, C$_{60}$ moves directly toward the cold end of the
graphene sheet when a temperature gradient is applied along armchair
direction. It is found that the lateral motion (along zigzag
direction) is a diffusive motion. The reduction of the free energy
of the system along the molecule motion is an indicative of the
drifted motion. Comparing the free energy difference with the work
performed on C$_{60}$  shows that the process is indeed
thermodynamically irreversible. The proposed mechanism for driving
nanoparticles on a graphene sheet may  be used in the design of
novel nanoscale motors.
\section{Acknowledgment}
 We gratefully acknowledge valuable comments from Hamid Reza
 Sepangi and Ali Naji.

\end{document}